\documentstyle[12pt,epsfig,dina4]{article}
\begin{document}
\baselineskip=15pt

\begin{center}

{\Large\bf
Breakup Density in Spectator Fragmentation
}
\vspace*{1.0cm}

\noindent
{
S.~Fritz,$^{(1)}$
C.~Schwarz,$^{(1)}$
R.~Bassini,$^{(2)}$
M.~Begemann-Blaich,$^{(1)}$
S.J.~Gaff-Ejakov,$^{(3)}$
D.~Gourio,$^{(1)}$
C.~Gro\ss,$^{(1)}$
G.~Imm\'{e},$^{(4)}$
I.~Iori,$^{(2)}$
U.~Kleinevo\ss,$^{(1)}$\cite{AAA}
G.J.~Kunde,$^{(3)}$\cite{BBB}
W.D.~Kunze,$^{(1)}$
U.~Lynen,$^{(1)}$
V.~Maddalena,$^{(4)}$			
M.~Mahi,$^{(1)}$
T.~M\"ohlenkamp,$^{(5)}$
A.~Moroni,$^{(2)}$
W.F.J.~M\"uller,$^{(1)}$
C.~Nociforo,$^{(4)}$       		
B.~Ocker,$^{(6)}$
T.~Odeh,$^{(1)}$\cite{AAA}
F.~Petruzzelli,$^{(2)}$
J.~Pochodzalla,$^{(1)}$\cite{CCC}
G.~Raciti,$^{(4)}$
G.~Riccobene,$^{(4)}$			
F.P.~Romano,$^{(4)}$    		
A.~Saija,$^{(4)}$     			
M.~Schnittker,$^{(1)}$
A.~Sch\"uttauf,$^{(6)}$\cite{DDD}
W.~Seidel,$^{(5)}$
V.~Serfling,$^{(1)}$
C.~Sfienti,$^{(4)}$			
W.~Trautmann,$^{(1)}$
A.~Trzcinski,$^{(7)}$
G.~Verde,$^{(4)}$
A.~W\"orner,$^{(1)}$
Hongfei~Xi,$^{(1)}$
and B.~Zwieglinski$^{(7)}$
}

\vspace*{0.5cm}

\noindent
{\it
$^{(1)}$Gesellschaft  f\"ur  Schwerionenforschung, D-64291 Darmstadt, 
Germany\\
$^{(2)}$Istituto di Scienze Fisiche, Universit\`{a} degli Studi
di Milano and I.N.F.N., I-20133 Milano, Italy\\
$^{(3)}$Department of Physics and
Astronomy and National Superconducting Cyclotron Laboratory,
Michigan State University, East Lansing, MI 48824, USA\\
$^{(4)}$Dipartimento di Fisica dell' Universit\`{a}
and I.N.F.N.,
I-95129 Catania, Italy\\
$^{(5)}$Forschungszentrum Rossendorf, D-01314 Dresden, Germany\\
$^{(6)}$Institut f\"ur Kernphysik,
Universit\"at Frankfurt, D-60486 Frankfurt, Germany\\
$^{(7)}$Soltan Institute for Nuclear Studies,
00-681 Warsaw, Hoza 69, Poland
}

\vspace{0.3cm}

\end{center}

{\bf
ABSTRACT}
\vspace{0.3cm}

Proton-proton correlations and correlations of protons, deuterons and 
tritons with $\alpha$~particles from spectator decays following
$^{197}$Au + $^{197}$Au collisions at 1000 MeV per nucleon have been 
measured with two highly efficient detector hodoscopes.
The constructed correlation functions, interpreted within the 
approximation of a simultaneous volume decay, indicate a moderate 
expansion and low breakup densities, similar to assumptions made 
in statistical multifragmentation models.
\vspace{2cm}


\newpage

Expansion is a rather basic conceptual feature of the 
multifragmentation of heavy nuclei. 
A volume of about three to eight times that occupied at saturation density
is assumed in the statistical models aiming at a phase space description
of the multi-fragment breakups \cite{gross90,bond95}.
Expansion also provides the link to the nuclear liquid-gas phase 
transition; only one third of the saturation value is expected for
the critical density of nuclear matter \cite{jaqa83}. 
The experimental confirmation of expansion 
to low breakup densities is therefore 
of the highest significance for the understanding and interpretation of the 
multifragmentation phenomenon \cite{more93}.

In central collisions of heavy nuclei, rapid expansion is evident from the
observation of radial collective flow \cite{reis97a}.
For the fragment decay of spectators following 
collisions at relativistic energies,
the case studied in this work,
significant radial flow has not been observed \cite{schuetti}.
Here evidence for expansion has been obtained, 
indirectly, from model comparisons.
Models that assume sequential emission from 
nuclear systems at saturation density underpredict 
the fragment multiplicities while those assuming expanded breakup
volumes yield satisfactory descriptions 
of the populated partition space \cite{bowm91,hubel92,hagel94,botv95}.
The disappearance of the Coulomb peaks in the kinetic-energy spectra of 
emitted light particles and fragments, associated with increasing fragment 
production, provides additional evidence consistent with volume 
emission from expanded systems \cite{milkau91,kwiat95}. There are also
other dynamical and statistical observables that have been interpreted as 
evidence for expansion in recent papers \cite{haug98,bege98,avde98}.

Interferometry-type methods permit experimental determinations of the 
breakup volume or, more precisely, of the space-time locations of the
last collisions of the emitted products \cite{ardouin}. 
In the nuclear regime, correlation functions for light charged 
particles, predominantly proton-proton correlations, have been widely 
explored for that purpose \cite{poch87,gong91}. 
Depending on the assumed reaction scenario and energy regime,
both time scales and breakup radii have been deduced. 
The time scales for the decay of highly excited spectator nuclei
produced at relativistic bombarding 
energies should be rather short \cite{wang98}, 
and we may expect that the correlation functions are mainly sensitive to 
the spatial extension of the source. More importantly, if we assume
a rapid volume breakup of the system, the quantity of interest will be
the local density, i.e. the mutual proximity of the nascent fragments 
and light particles. These densities are obtained in the limit of
assuming a zero-lifetime in the source analysis.

In this Letter, we present the results of correlation measurements for 
spectator decays following collisions of 
$^{197}$Au + $^{197}$Au at a bombarding energy of 1000 MeV per nucleon.
Besides proton-proton coincidences, also coincidences of protons,
deuterons, and tritons with $\alpha$ particles were measured and
correlation functions were constructed. 
For their quantitative interpretation, it is assumed that they 
are dominated by the effect of final-state interactions. 
The results are found to be consistent with low breakup densities 
with values close to those assumed in the statistical multifragmentation 
models. 

Beams of $^{197}$Au with incident energy 1000 MeV per nucleon were 
provided by the heavy-ion synchrotron SIS and directed onto targets 
of 25-mg/cm$^2$ areal thickness.
Two multi-detector hodoscopes, consisting of a total of 
160 Si-CsI(Tl) telescopes in closely-packed geometry, were placed
on opposite sides with respect to the beam axis. The angular range
$\theta_{lab}$ from 122$^{\circ}$ to 156$^{\circ}$ was chosen with the
aim of selectively detecting the products of the target-spectator decay.
Each telescope consisted
of a 300-$\mu$m Si detector with 30 x 30 mm$^2$ (96 detectors) or
25 x 25 mm$^2$ (64 detectors) active area, followed
by a 6-cm long CsI(Tl) scintillator with photodiode readout. 
The distance to the target was about 60 cm.

The products of the projectile decay were measured with the time-of-flight 
wall of the ALADIN spectrometer \cite{schuetti} and the quantity $Z_{bound}$
was determined event-by-event. 
$Z_{bound}$ is defined as the sum of the atomic numbers
$Z_i$ of all projectile fragments with $Z_i \geq$ 2. It reflects 
the variation of the size of the primary
spectator nuclei and is inversely correlated with its excitation energy. 
Because of the
symmetry of the collision system, the mean values of $Z_{bound}$
for the target and the projectile spectators 
within the same event class have been assumed to be identical.

\begin{figure}[tb]
  \centerline{\epsfig{file=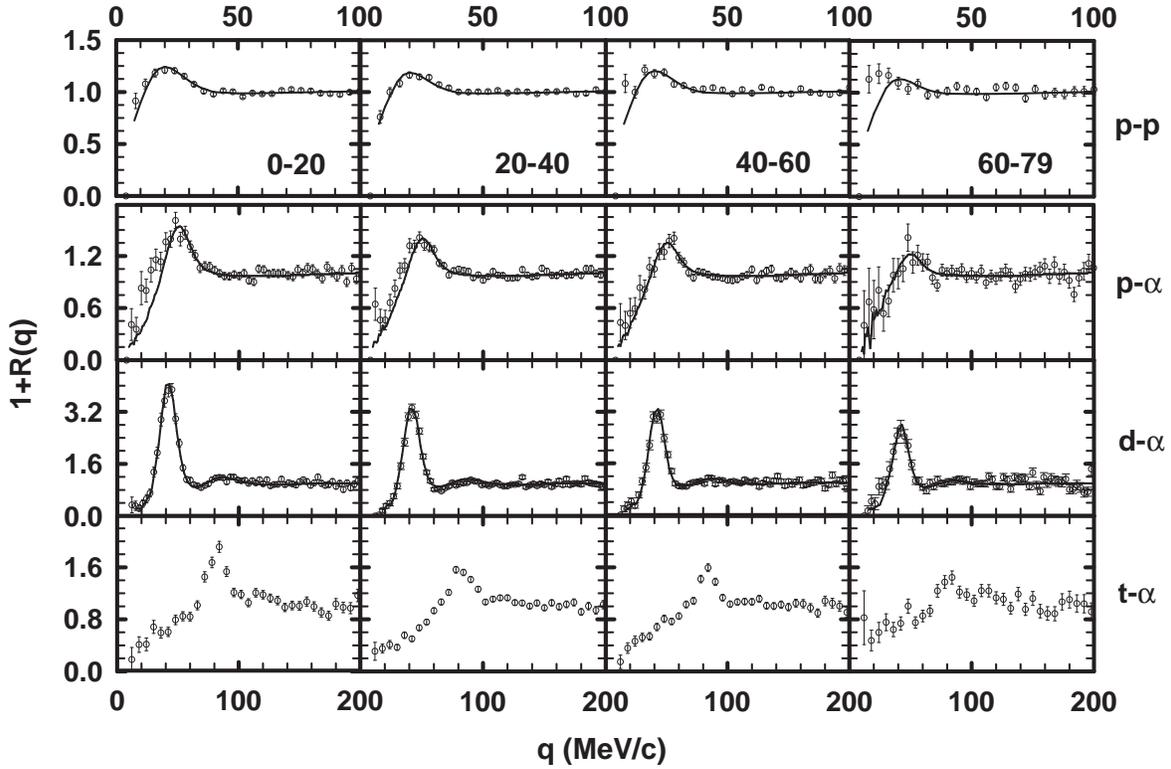,width=0.95\textwidth}}
\caption[]{
  Correlation functions constructed for p-p, p-$^4$He, d-$^4$He, and
  t-$^4$He coincidences (from top to bottom) for spectator decays following
  $^{197}$Au + $^{197}$Au collisions at 1000 MeV per nucleon. The data are
  sorted according to $Z_{bound}$ into four bins with limits given 
  in the top panel of each column. 
  The lines represent the 
  results of the calculations used to extract source radii (see text).
  Note that the scales of the 
  abscissa are different for the top row (p-p correlations) and for 
  the remaining three rows of panels.
}
\end{figure}

Examples of correlation functions constructed for the four types of 
coincidences
p-p, p-$^4$He, d-$^4$He, and t-$^4$He are shown in Fig. 1, sorted into
four bins of $Z_{bound}$ as indicated. The uncorrelated yields were 
obtained with the technique of event-mixing and normalized 
in the range of large relative momenta $q \ge$ 70 MeV/c
(p-p) and $q \ge$ 150 MeV/c (p-$\alpha$, d-$\alpha$, and t-$\alpha$).
In the off-line analysis, thresholds were set at 
$E_{lab} \ge$ 20 MeV for all particles p, d, t, and $\alpha$ 
and all hodoscope detectors. The shapes of the correlation functions are 
sensitive to the threshold. For p-p, e.g., the suppression at small $q$
tends to disappear if the threshold is set at $E_{lab} \ge$ 10 MeV or lower,
presumably as a result of increasing contributions from evaporation and
sequential decay. These long-lifetime components are suppressed if higher 
thresholds are chosen. The pairs of particles were also requested to be
detected in the same hodoscope in order to avoid correlation effects
that appear at large $q$ and are believed to be due to
a collective (sidewards) motion of the proton-emitting source.

The p-p correlation functions are characterized by a depression 
at small relative momentum and by a weakly pronounced 
peak near relative momentum $q$ = 20 MeV/c, caused by 
the S-wave nuclear interaction and used for the quantitative interpretation.
The three correlation functions of the hydrogen isotopes with 
$\alpha$ particles are
dominated by the resonances corresponding
to the ground state of $^5$Li and by the 2.19-MeV and 4.63-MeV
excited states of $^6$Li and $^7$Li, respectively. The observed widths of the
$^6$Li and $^7$Li peaks in the d-$\alpha$ and t-$\alpha$ correlation 
functions represent the experimental resolution which is mainly 
determined by the angular resolution following from the geometry
of the detector hodoscopes.
A striking feature of the data is the overall stability of the
peak heights as a function of $Z_{bound}$ which indicates 
source extensions that do not change dramatically with impact parameter.
The largest deviations of the peak height and of the overall shape of the
correlation functions appear in the bin 60 $\le Z_{bound}\le$ 79, 
corresponding to the largest impact parameters. 

The analysis of the p-p correlation functions
was performed with the Koonin-Pratt formalism \cite{pratt87} in the
zero-lifetime limit. In this form,
the analysis includes the effects of quantum statistics and
of the mutual nuclear and Coulomb final-state interactions but it 
ignores the long-range Coulomb repulsion of the two protons
from the emitting source. In order to assess the magnitude of the 
latter effect, classical Coulomb trajectory calculations were performed. 
In addition, calculations with the three-body quantum model of 
Lednicky {\it et al.} \cite{mart96} were used to identify possible 
systematic uncertainties.
A uniform sphere with radius $R$ and statistical momentum distributions 
corresponding to the measured kinetic-energy spectra 
were assumed for the proton source; the model results are slightly
dependent on the particle momenta through the applied experimental filter.

\begin{figure}[tb]
\begin{minipage}[t]{0.5\textwidth}
       \centerline{\epsfig{file=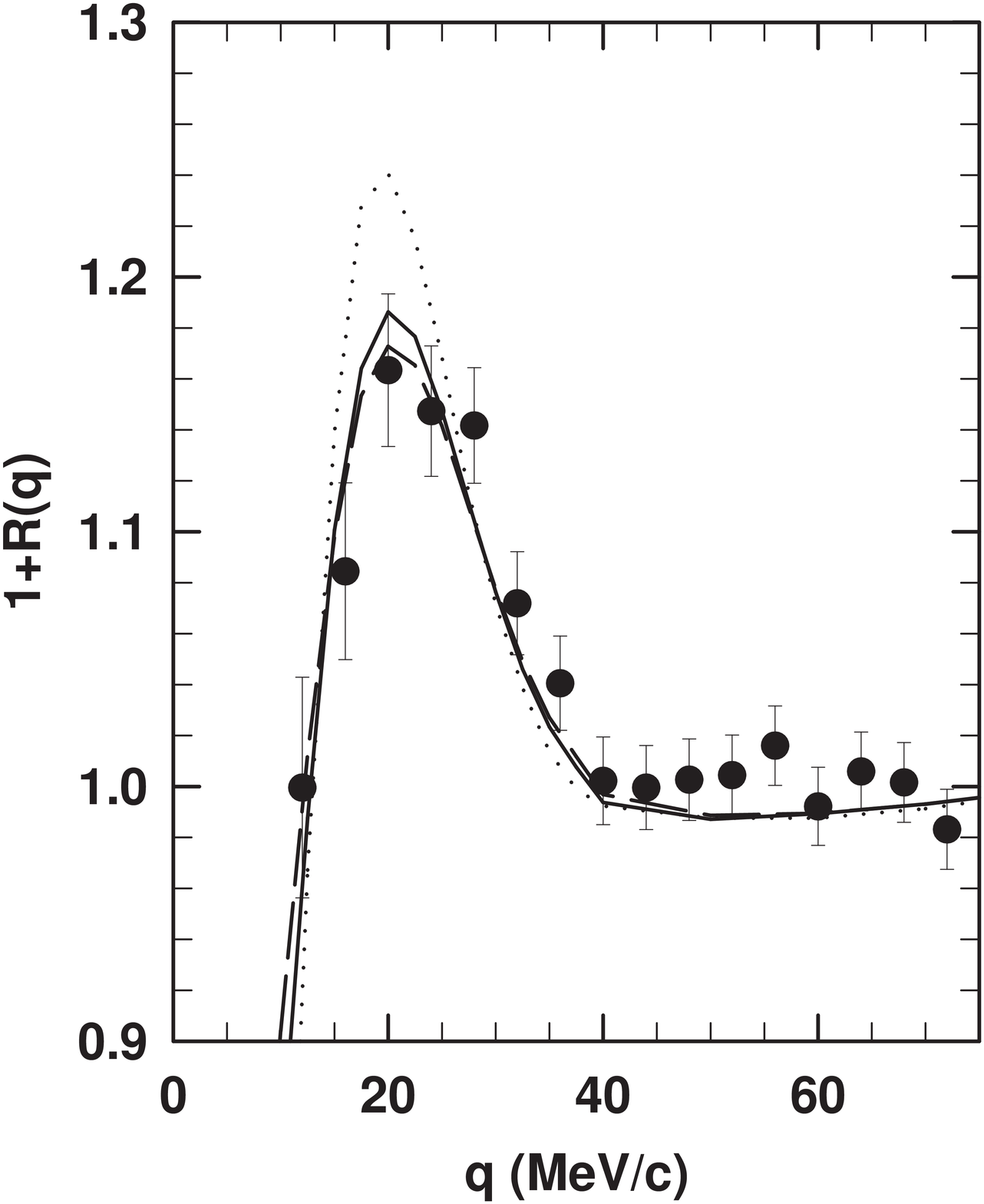,width=1.0\textwidth}}
\end{minipage}
\hfill
\parbox[b]{0.47\textwidth}{
\caption[]{
  Proton-proton correlation function for the interval 20 $\le Z_{bound} \le$
  40 in comparison with the predictions of the Koonin-Pratt formalism
  before (dotted) and after applying the experimental filter (full line),
  and after correction for the Coulomb repulsion from the emitting source
  (dashed line, closely following the full line). 
  The calculations were
  performed for a homogeneous spherical source with radius $R$ = 8.2 fm.
}
}
\end{figure}

The quality of the obtained results and
the sensitivity to the radius parameter $R$
is illustrated in Fig. 2 for the case 20 $\le Z_{bound} \le$ 40.
With the filtered Koonin-Pratt calculations, the most satisfactory 
description of the data is obtained with $R$ = 8.2 fm (full line). 
The height and the width of the resonance peak are rather well reproduced.
For comparison, the results of the same calculation before filtering
(dotted line) and after correction for the Coulomb repulsion from the
source (dashed line) are also shown.
The filtering effect is equivalent to a change of the radius by
$\Delta R \approx$ 0.3 fm whereas the modification caused by the 
charge of the emitting source is almost negligible.
The latter is not unexpected in the present case, 
especially for the more central 
collisions which produce spectators of very moderate total charge
(e.g. $\approx$ 45 for 20~$\le Z_{bound} \le$~40).
In addition, with the assumption of a simultaneous volume 
break-up, only the charge inside the volume corresponding to the radial
position of the particle was assumed to contribute to its acceleration.
Correlation peaks of nearly the same height but with slightly smaller width 
were obtained with the three-body quantum model
\cite{mart96}. 
In the two-body approximation, the difference between the results 
obtained with the Koonin-Pratt and the Lednicky {\it et al.} formalisms 
is practically negligible. Therefore,
and because of the reduced importance of the three-body Coulomb effect,
the quantitative analysis of the measured p-p correlation functions was
performed with the filtered Koonin-Pratt simulations.

\begin{table}[bt]

\begin{center}

\begin{tabular}{|l|c|c|c|c|}
\hline
$Z_{bound}$ & 0-20 & 20-40 & 40-60 & 60-79 \\
\hline
$<A>$ & 50 & 110 & 150 & 184 \\
\hline
p-p & 
 7.8  $\pm$ 0.1  & 
 8.2  $\pm$ 0.2  & 
 8.1  $\pm$ 0.2  & 
 8.8  $\pm$ 0.6  \\ 
   & (0.7) & (1.4) & (1.0) & (3.2) \\ 
\hline
p-$\alpha$ &
 9.5  $\pm$ 0.4  &
10.5  $\pm$ 0.4  &
10.9  $\pm$ 0.3  &
12.7  $\pm$ 0.8   \\
   & (3.3) & (2.6) & (0.8) & (0.8) \\
\hline
d-$\alpha$ &
 8.4  $\pm$ 0.2  &
 9.3  $\pm$ 0.1  &
 9.5  $\pm$ 0.2  &
10.5  $\pm$ 0.2   \\
   & (0.7) & (1.2) & (1.6) & (0.3) \\
\hline
\end{tabular}

\end{center}

\caption{Mean mass numbers $<A>$ and 
extracted source radii in units of fm for the indicated bins of
$Z_{bound}$. The regions of relative momentum $q$ selected for the 
$\chi^2$ test are 10 - 35, 30 - 70, and 30 - 54 MeV/c for the 
p-p, p-$\alpha$, and d-$\alpha$ correlation functions, respectively.
The numbers in brackets denote the minimum $\chi^2$ per degree of 
freedom.}   
\end{table}

The best fits, generated by minimizing $\chi ^2$ within the range 
10~$\le q \le$~35~MeV/c are shown in Fig. 1. 
The corresponding radii, listed in Table 1, are close to about 8 fm, 
up to nearly 9 fm for the bin of largest $Z_{bound}$ which, however, 
seems to be afflicted with the largest uncertainties. 
These values are distinctly 
but not excessively larger than the radius $R$ = 6.7~fm 
of a gold nucleus at normal density 0.16~fm$^{-3}$.
The quoted errors are purely statistical and were determined from the
radii for which $\chi^2$ exceeds its minimum by an amount equal to
the minimum value of $\chi^2$ per degree of freedom
(i.e. for $\chi^2 = \chi^2_{min} \cdot n/(n-1)$
where $n$ is the number of data points included).
The additional systematic uncertainty, mainly resulting from the
arbitrariness of choosing the normalization interval and
estimated to be about 0.5 fm, is much larger.

\begin{figure}[tb]
\begin{minipage}[t]{0.6\textwidth}
  \centerline{\epsfig{file=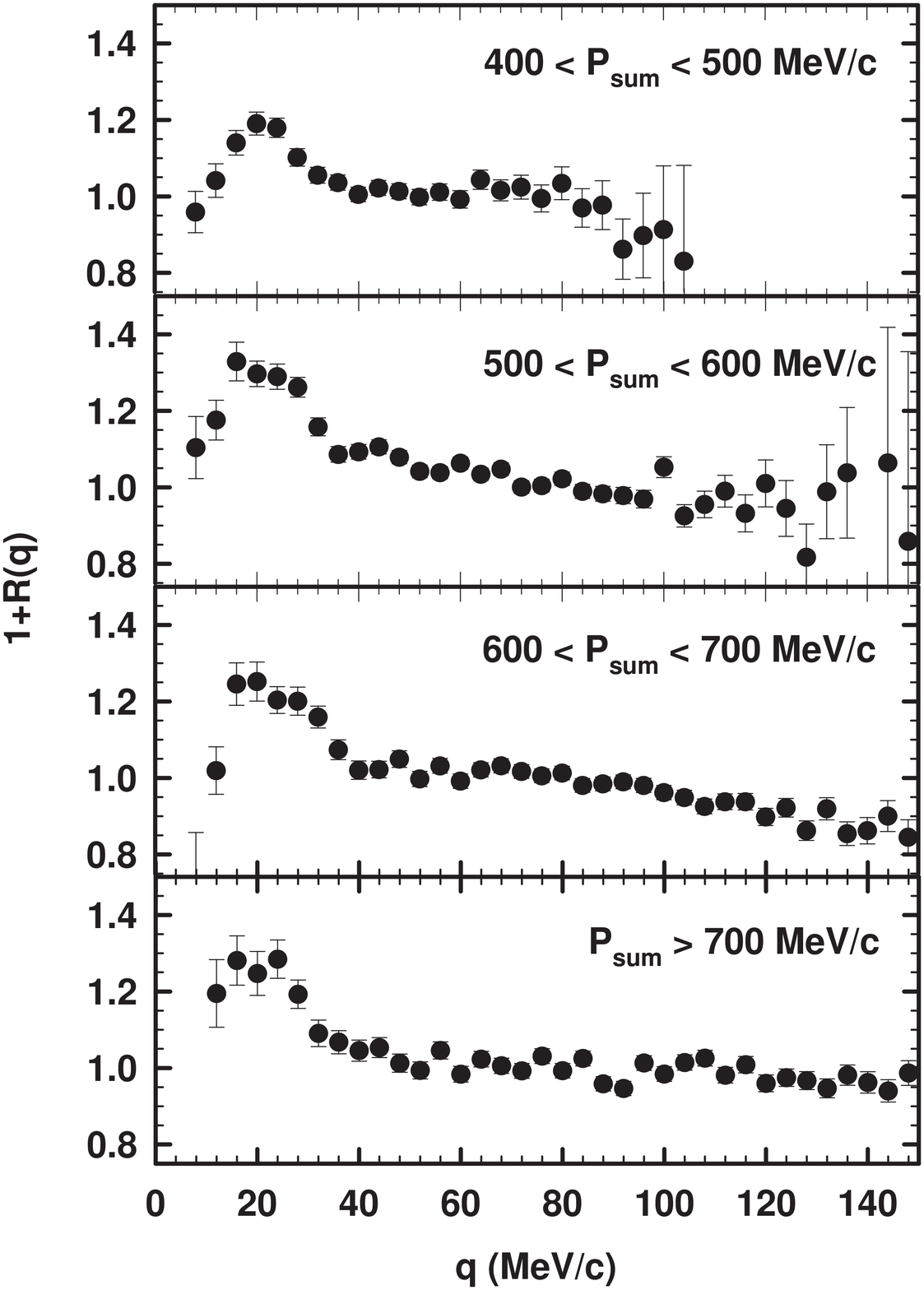,width=1.0\textwidth}}
\end{minipage}
\hfill
\parbox[b]{0.37\textwidth}{
\caption[]{
  Proton-proton correlation functions, integrated over $Z_{bound}$, for 
  four intervals of the absolute value of the proton pair momentum. 
  The interval 70 $\le q \le$ 90 MeV/c was used for normalization.
}
}
\end{figure}

Correlation functions for proton pairs belonging to four intervals 
of the laboratory pair momentum $P_{sum} = |\vec p_1 + \vec p_2|$,
integrated over $Z_{bound}$, are shown in Fig. 3. 
Only the data from the 96-element hodoscope were used.
The combined effects of the energy threshold $E_{lab} \ge$ 20 MeV, 
of the binning in pair momentum, and of the finite solid-angle acceptance 
of the detector hodoscope cause the limitation of the populated $q$ range 
for the bins of smaller $P_{sum}$. 
A normalization interval 70 $\le q \le$ 90 MeV/c
was therefore chosen for these correlation functions. The same effects,
perhaps including some collectivity due to a small but finite source 
motion, may cause the slight decrease towards
larger $q$ that is not observed in the momentum-integrated correlation
functions (cf. Fig. 1). We observe, however, that the peak height at 
$q \approx$ 20 MeV/c relative to the
uncorrelated background at $q >$ 40 MeV/c is virtually identical in all four 
cases. This is consistent with what is expected for the ideal situation of 
a purely statistical source, confirming that effects caused 
by collective radial motion or emission times related to the temporal
evolution of an expanding source
should be small for spectator decays.

The choice of a threshold of $E_{lab} \ge$ 20 MeV, on the other hand, raises
the question of what reaction stage is mainly
represented by the obtained correlation data. 
Light-particle spectra from the same experiment, 
measured with high-resolution telescopes at backward angles, 
give evidence for emission prior to the final breakup stage \cite{hongfei}. 
This was concluded from the comparison of the 
slope temperatures and multiplicities to the predictions of the statistical 
multifragmentation model. Distinct components of equilibrium yields 
and of faster particles, termed pre-equilibrium or first stage, 
have also been identified by other groups in data for comparable 
reactions \cite{haug98,kwiat98}. In all these cases the main parts
of the low-energy equilibrium components are below the 
present threshold. Long-lifetime components are thus excluded 
from the analysis so as to preserve the sensitivity of the 
correlation function to the initial spatial
dimensions. On the other hand, pre-equilibrium or pre-breakup particles 
as they are scattered from the forming spectator matter may also
contribute to an
interferometric picture that reflects the extension of the latter.

Besides the p-p correlations also the p-$\alpha$ and d-$\alpha$
correlations were used to determine breakup radii by comparing them to the
numerical results of Boal and Shillcock \cite{boal86}. The calculated 
correlation functions which in their work are given 
for a discrete set of source radii were interpolated, and a
Monte Carlo procedure was used for an event-by-event simulation of
the effects of multiple scattering in the target and of the spatial 
and energy resolution of the detection system \cite{carsten}. 
The Coulomb acceleration by the emitting source was also here neglected.
This seems justified for d-$\alpha$ because the effects 
should not be much larger than in the p-p case. For p-$\alpha$, the different
charge-to-mass ratios of the two particles may be expected to cause 
larger distortions but, at the same time, the finite lifetimes of these 
resonances may reduce them considerably.
The resulting fits to the data are rather satisfactory (Fig. 1). 
The corresponding radii are about 1 to 2 fm larger than those 
deduced from the p-p correlations (Table 1). We also observe a slightly
stronger variation with the pair momentum than in the p-p case. 
For p-$\alpha$, e.g., the peak of the resonance grows from about 1.3 to 
1.6 as the pair momentum is varied from $P_{sum} \le$ 800 MeV/c to
$P_{sum} >$ 1000 MeV/c (result after integration over $Z_{bound}$). 

\begin{figure}[tb]
  \centerline{\epsfig{file=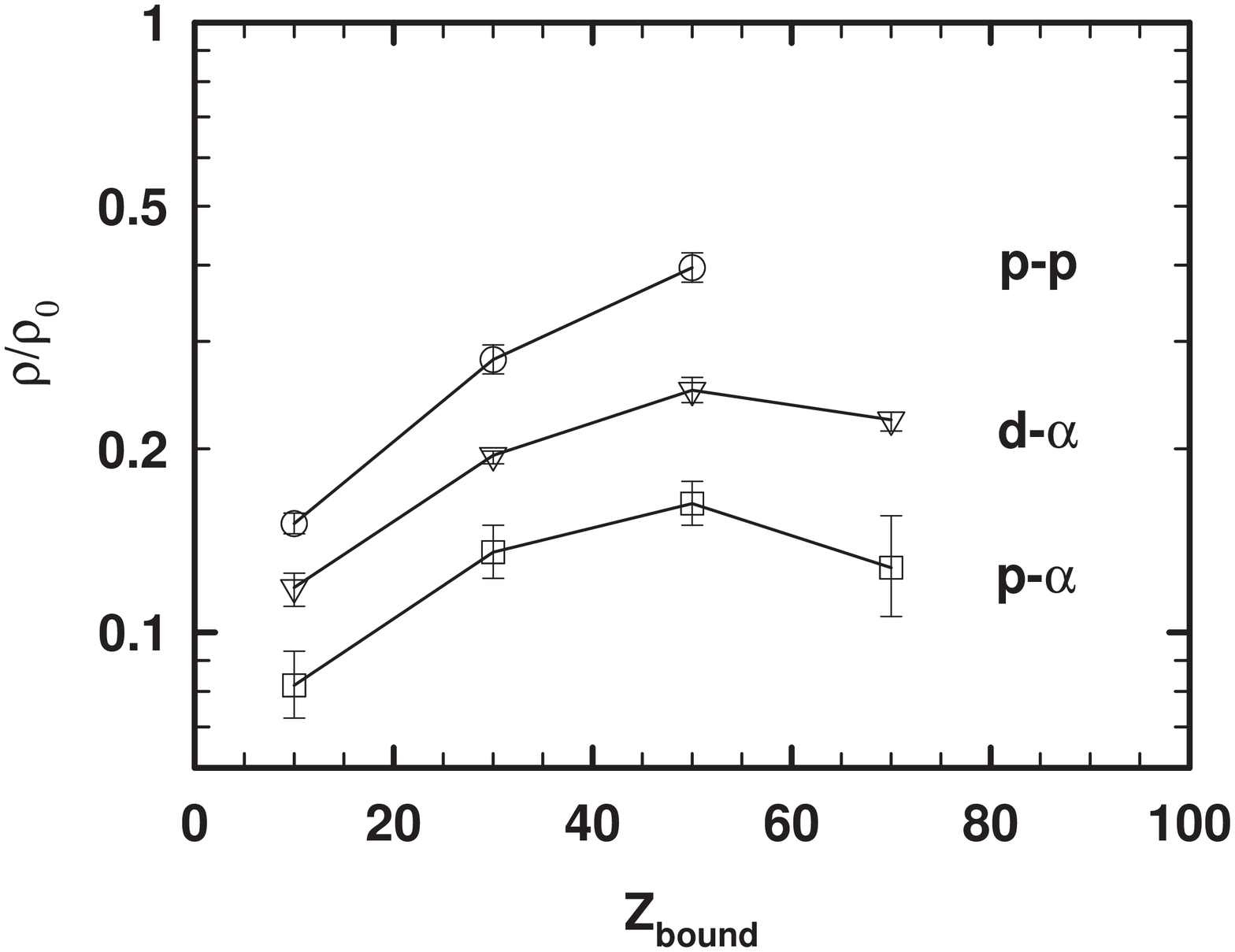,width=0.8\textwidth}}
\caption[]{
  Breakup density $\rho /\rho_0$ as deduced from the 
  p-p, p-$^4$He, and d-$^4$He correlation functions. The rapid 
  decrease of the density with decreasing $Z_{bound}$ reflects the changing 
  mass of the spectator source. The error bars represent the statistical 
  uncertainty of the radii as given in the table (see text). 
  Note the logarithmic ordinate scale.
}
\end{figure}

Densities were calculated by dividing the number of spectator 
constituents, taken from the calorimetric
analyses of Refs. \cite{poch95,gross98} and listed in Table~1,
by the source volume. The  densities vary considerably with centrality 
(Fig. 4) even though the radii are approximately constant.
This is caused by the varying spectator masses which, in excellent 
agreement with the
prediction of the geometric participant-spectator model \cite{gosset},
decrease with increasing
centrality almost in proportion to $Z_{bound}$.

In the p-p case, the mean relative densities decrease 
from $\rho /\rho_0$ = 0.4 for the near-peripheral to below 
$\rho /\rho_0$ = 0.2 for the most central collisions.
These values compare well with the densities
assumed in the statistical multifragmentation model, including their 
variation with centrality. In the model, the mean density changes as 
a function of the multiplicity if a fixed socalled crack width is used as 
criterion for the placement of fragments 
inside the breakup volume \cite{bond95}.
For the most peripheral bin, the interpretation of the shape of the
measured p-p correlations is somewhat uncertain (cf. Fig. 1),
and the corresponding density has not been plotted. 
For p-$\alpha$ and d-$\alpha$, the fits are satisfactory
for the largest $Z_{bound}$ 
but correspond to rather large radii (Table 1). The densities are
therefore low which is somewhat unexpected as here
the production of highly excited heavy residues is the dominant reaction 
channel \cite{schuetti}. Obviously, the assumptions of a 
homogeneous spherical source and of a rapid volume breakup are less well
justified in this case which causes difficulties for the interpretation of
the obtained density values.

The differences between the results for p-p and for the 
resonances involving $\alpha$ particles are statistically significant but
their origin is not clear at present. 
Besides the systematic uncertainties
of the employed formalisms, there is also the possibility of 
differences in the scattering cross sections; larger cross 
sections will cause larger apparent source sizes. An interesting
connection exists between the large radii derived for the lithium
states and the limitation of the temperatures 
deduced from their relative populations \cite{serf98}. A late emission 
of these resonances could explain both observations.

In summary, correlation functions constructed from proton-proton
coincidences and from coincidences between protons, deuterons, or tritons
with $\alpha$ particles consistently show that the breakup volume
does not appreciably change with impact parameter even though the spectator 
mass varies considerably. A quantitative analysis, in the limit of
zero lifetime of the source, yields results
that are consistent with a very moderate radial expansion. The deduced
breakup densities are rather low 
as assumed in the statistical model scenarios. 
The variation of the density with impact parameter is caused 
by the changing spectator mass. 
With regard to the method, it has become clear
that the systematic uncertainties of deriving densities
from correlation functions may be large, in particular for the resonances
involving $\alpha$ particles for which the formalism is not yet as 
much advanced.
It seems, however, that the spectator decay at 
relativistic energies may represent a particularly favorable case
because the source charge is
moderate, collective motion is nearly nonexistent, and with the choice of
high energy thresholds long-lifetime components may have been
efficiently excluded from the analysis.

The authors would like to thank R.~Lednicky for making his code 
available to them and W.A.~Friedman for fruitful discussions. 
M.B., J.P., and C.S. acknowledge the financial support
of the Deutsche Forschungsgemeinschaft under the Contract No. Be1634/1-1,
Po256/2-1, and Schw510/2-1, respectively.
This work was supported by the European Community under
contract ERBFMGECT950083.

\end{document}